\documentstyle[sprocl]{article}

\def\be{\begin{equation}}
\def\ee{\end{equation}}
\def\bea{\begin{eqnarray}}
\def\eea{\end{eqnarray}}

\input{tcilatex}

\begin{document}

\title{PATH INTEGRAL ON RELATIVISTIC SPINLESS POTENTIAL PROBLEMS}
\author{De-Hone Lin}
\address{Department of Physics, National Tsing Hua University\\ 
Hsinchu 30043, Taiwan \\
E-mail: d793314@phys.nthu.edu.tw}
\maketitle
\abstracts{\ The formulation of the relativistic spinless path integral on
the general affine space is presented. For the one dimensional space, the
Duru-Kleinert (DK) method and the $\delta $-function perturbation technique
are applied to solve the relativistic path integrals of the smooth potential
and the Dirichlet boundary condition problems, respectively.}

\section{Relativistic Path Integral on General Affine Space}

Let us first consider a point particle of mass $M$ moving at a relativistic
velocity in a $(D+1)$-dimensional Minkowski space with a given
electromagnetic field. By using $t=-i\tau =-ix^{4}$, the path integral
representation of the fixed-energy amplitude is conveniently formulated in a 
$(D+1)$-Euclidean spacetime with the Euclidean metric, 
\begin{equation}
(g_{\mu \nu })={\rm diag}\;\,(1,\cdots ,1,c^{2}),  \label{2.1}
\end{equation}
and is given by \cite{1,3} 
\begin{equation}
G({\bf x}_{b},{\bf x}_{a};E)=\frac{i\hbar }{2Mc}\int_{0}^{\infty }dL\int 
{\cal D}\rho (\lambda )\Phi \left[ \rho (\lambda )\right] \int {\cal D}%
^{D}x(\lambda )e^{-A_{E}/\hbar }  \label{2.2}
\end{equation}
with the action 
\begin{equation}
A_{E}=\int_{\lambda _{a}}^{\lambda _{b}}d\lambda \left[ \frac{M}{2\rho
\left( \lambda \right) }{\bf x}^{\prime ^{2}}\left( \lambda \right) -i\frac{e%
}{c}{\bf A(x)\cdot x^{\prime }(}\lambda {\bf )}-\rho (\lambda )\frac{\left(
E-V({\bf x})\right) ^{2}}{2Mc^{2}}+\rho \left( \lambda \right) \frac{Mc^{2}}{%
2}\right] ,  \label{2.3}
\end{equation}
where $L$ is defined by 
\begin{equation}
L=\int_{\lambda _{a}}^{\lambda _{b}}d\lambda \rho (\lambda ),  \label{2.4}
\end{equation}
in which $\rho (\lambda )$ is an arbitrary dimensionless fluctuating scale
variable, and $\Phi \lbrack \rho ]$ is some convenient gauge-fixing
functional, such as $\Phi \left[ \rho \right] =\delta \left[ \rho -1\right] $%
, to fix the value of $\rho (\lambda )$ to unity. $\hbar /Mc$ is the
well-known Compton wave length of a particle of mass $M$, ${\bf A(x)}$ is
the vector potential, $V({\bf x})$ is the scalar potential, $E$ is the
system energy, and ${\bf {x}}$ is the spatial part of the ($D+1$) vector $x=(%
{\bf {x}},\tau )$. This path integral forms the basis for studying
relativistic potential problems. To obtain a tractable path integral in (\ref
{2.2}) for the arbitrary time independent potential $V({\bf x})$, Let us
first take the space-time transformation as usual \cite{2} 
\begin{equation}
\triangle \lambda _{n}=\triangle s_{n}f({\bf {x}}_{n}),  \label{2.5}
\end{equation}
where $\triangle \lambda _{n}\equiv \lambda _{n}-\lambda _{n-1}$ is a small
parameter interval and $f({\bf {x}})$ are invertible but otherwise arbitrary
function. A general formulation of relativistic path integral in affine
space is given by introducing the coordinate transformation 
\begin{equation}
x^{i}=h^{i}(q).  \label{2.11}
\end{equation}
The differential mapping may be written as 
\begin{equation}
dx^{i}=\partial _{\mu }h^{i}(q)dq^{\mu }=e_{~~\mu }^{i}(q)dq^{\mu }.
\label{2.12}
\end{equation}
Since we must find all terms that will eventually contribute to order $%
\epsilon $, we expand $(\triangle {\bf x}_{n})^{2}=({\bf x}_{n}-{\bf x}%
_{n-1})^{2}$ up to fourth order in $\triangle q_{n}^{\mu }=q_{n}^{\mu
}-q_{n-1}^{\mu }$. Finally, we obtain the path integral representation for
the fixed-energy amplitude of a relativistic particle in general affine
space \cite{1,2,3}

\[
G({\bf x}_{b},{\bf x}_{a};E)=\lim_{n\rightarrow \infty }\frac{i\hbar }{2Mc}%
\int_{0}^{\infty }dS\frac{f(q_{a})}{\left( \frac{2\pi \hbar \epsilon
_{b}^{s}\rho _{b}f(q_{a})}{M}\right) ^{D/2}}\prod_{n=1}^{N+1}\left[ \int
d\rho _{n}\Phi (\rho _{n})\right] 
\]
\begin{equation}
\times \prod_{n=2}^{N+1}\left[ \int_{-\infty }^{\infty }\frac{%
d^{D}\bigtriangleup q_{n}g^{1/2}(q_{n})}{\left( \frac{2\pi \hbar \epsilon
_{n}^{s}\rho _{n}f(q_{n})}{M}\right) ^{D/2}}\right] \exp \left\{ -\frac{1}{%
\hbar }\sum_{n=1}^{N+1}\left[ A^{\epsilon }+A_{J}^{\epsilon }+A_{{\rm pot}%
}^{\epsilon }\right] \right\}  \label{2.30}
\end{equation}
with each $s$-sliced action 
\[
A^{\epsilon }+A_{J}^{\epsilon }+A_{{\rm pot}}^{\epsilon }=\frac{M}{2\epsilon
^{s}\rho f}g_{\mu \nu }\triangle q^{\mu }\triangle q^{\nu }-\frac{\hbar }{2}%
\Gamma _{\mu ~~~\nu }^{~~\mu }\triangle q^{\nu }+\epsilon ^{s}\rho f\frac{%
\hbar ^{2}}{8M}(\Gamma _{\mu ~~~\nu }^{~~\mu })^{2} 
\]
\begin{equation}
-i\frac{e}{c}\left[ A_{\mu }\triangle q^{\mu }-\epsilon ^{s}\rho f\frac{%
\hbar }{2M}\left( A_{\nu }\Gamma _{\mu }^{\;\mu \nu }+D_{\mu }A^{\mu
}\right) \right] -\epsilon ^{s}\rho f\frac{\left( E-V(q)\right) }{2Mc^{2}}%
^{2},  \label{2.31}
\end{equation}
where we have omitted the subscripts $n$ of $\triangle q^{\mu }$. The
induced metric $g_{\mu \nu }=e_{~\mu }^{i}e_{~\nu }^{i}$ and affine
connection $\Gamma _{\lambda \kappa }^{\quad \mu }=e_{i}^{~~\mu }e_{~\kappa
,\lambda }^{i}$ are evaluated at the postpoint, $\dot{q}^{\alpha }$ stands
for $dq^{\alpha }/ds.$ The action of covariant derivative $D_{\mu }$ is
defined as 
\begin{equation}
D_{\mu }A^{\mu }=\partial _{\mu }A^{\mu }+\Gamma _{\mu \lambda }^{\;\;\;\mu
}A^{\lambda }.  \label{2.32}
\end{equation}
Remarkably, this expression involves only the connection contracted in the
first two indices: 
\begin{equation}
\Gamma _{\mu }^{\;\;\,\,\mu \nu }=g^{\mu \lambda }\Gamma _{\mu \lambda
}^{\quad \nu }.  \label{2.33}
\end{equation}

\section{Duru-Kleinert Method for the Relativistic Potential Problems}

The relativistic stable path integrals of Eq. (\ref{2.2}) have a more
elegant representation if the systems are in two-dimensional Minkowski space
or rotationally invariant systems in any dimensions. By decomposing the Eq. (%
\ref{2.2}) into angular parts and take DK transformation, we get \cite{2} 
\begin{equation}
G({\bf {x}}_{b},{\bf {x}}_{a};E)=\frac{1}{(r_{b}r_{a})^{(D-1)/2}}%
\sum_{l=0}^{\infty }G_{l}^{{\rm DK}}(r_{b},r_{a};E)\sum_{{\bf \hat{m}}}Y_{l%
{\bf {\hat{m}}}}({\bf {\hat{x}}}_{b})Y_{l{\bf \hat{m}}}^{\ast }({\bf \hat{x}}%
_{a}),  \label{3.1}
\end{equation}
where superscript ${\rm DK}$ indicates that the system has been performed by
the DK-transformation, the functions $Y_{l{\bf \hat{m}}}({\bf \hat{x}})$ are
the $D$-dimensional hyperspherical harmonics and $G_{l}^{{\rm DK}%
}(r_{b},r_{a};E)$ is the purely radical transformed amplitude 
\begin{equation}
G_{l}^{{\rm DK}}(r_{b},r_{a};E)=\frac{\hbar i}{2Mc}%
f_{b}^{1/4}f_{a}^{1/4}G(q_{b,}q_{a};E).  \label{3.2}
\end{equation}
The amplitude $G(q_{b,}q_{a};E)$ is given by \cite{2} 
\begin{equation}
G(q_{b},q_{a};E)\equiv \int_{0}^{\infty }dS\int {\cal D}\rho (s)\Phi \left[
\rho (s)\right] \int {\cal D}q(s)e^{-A_{s}^{{\rm DK}}/\hbar }  \label{3.3}
\end{equation}
with 
\[
A_{s}^{{\rm DK}}=\int_{0}^{S}ds\left[ \frac{M}{2\rho (s)}\dot{q}^{2}(s)+\rho
(s)f(q(s))\right. \qquad \qquad \qquad \qquad 
\]
\begin{equation}
\left. \times \left( \frac{\hbar ^{2}}{2M}\frac{(l+D/2-1)^{2}-1/4}{%
r^{2}(q(s))}-\frac{\left[ E-V(r(q))\right] ^{2}}{2Mc^{2}}+\frac{Mc^{2}}{2}%
\right) +V{\rm _{eff}}\right] .  \label{3.4}
\end{equation}
The effective potential $V{\rm _{eff}}$ is given by 
\begin{equation}
V{\rm _{eff}}=-\frac{\rho (s)\hbar ^{2}}{M}\left[ \frac{1}{4}\frac{h^{\prime
\prime \prime }(q)}{h^{\prime }(q)}-\frac{3}{8}\left( \frac{h^{\prime \prime
}(q)}{h^{\prime }(q)}\right) ^{2}\right] .  \label{3.5}
\end{equation}
Here the transformation function $r=h(q)$ with 
\begin{equation}
h^{\prime 2}(q)=f(r).  \label{3.8}
\end{equation}

\section{Relativistic Path Integral With Dirichlet Boundary Conditions via
Sum Over Perturbation Series}

With the help of the $\delta $-function perturbation techniques, we obtain
the relativistic fixed-energy amplitude for an impenetrable wall appears at $%
x=a$ given by \cite{3} 
\begin{equation}
G^{({\rm {wall}})}(x_{b},x_{a};E)=G_{0}(x_{b},x_{a};E)-\frac{%
G_{0}(x_{b},a;E)G_{0}(a,x_{a};E)}{G_{0}(a,a;E)}.  \label{4.7}
\end{equation}
Furthermore, for the relativistic motion inside a box with boundaries at $x=a
$ and $x=b$ one obtains ($a<x<b,$Dirichlet-Dirichlet boundary conditions) 
\cite{3} 
\[
G({x}_{b},{x}_{a};E)=\frac{i\hbar }{2Mc}\int_{0}^{\infty }dL\int {\cal D}%
\rho (\lambda )\Phi \left[ \rho (\lambda )\right] 
\]
\begin{equation}
\times \int_{x_{a}}^{x_{b}}{\cal D}x(\lambda )\exp \left\{ -\frac{1}{\hbar }%
\int_{\lambda _{a}}^{\lambda _{b}}d\lambda \left[ \frac{M}{2\rho \left(
\lambda \right) }{x}^{\prime ^{2}}\left( \lambda \right) -\rho (\lambda )%
\frac{(E-V(x))^{2}}{2Mc^{2}}+\rho \left( \lambda \right) \frac{Mc^{2}}{2}%
\right] \right\} 
\end{equation}
\begin{equation}
=\frac{\left| 
\begin{array}{ccc}
G_{0}(x_{b},x_{a};E) & G_{0}(x_{b},b;E) & G_{0}(x_{b},a;E) \\ 
G_{0}(b,x_{a};E) & G_{0}(b,b;E) & G_{0}(b,a;E) \\ 
G_{0}(a,x_{a};E) & G_{0}(a,b;E) & G_{0}(a,a;E)
\end{array}
\right| }{\left| 
\begin{array}{cc}
G_{0}(b,b;E) & G_{0}(b,a;E) \\ 
G_{0}(a,b;E) & G_{0}(a,a;E)
\end{array}
\right| },  \label{4.8}
\end{equation}
where the notation $G_{0}$ stands for the non-perturbative relativistic
fixed-energy amplitude.

The methods present in this paper enable us to solve a large class of the
relativistic potential problems. It is our hope that our studies would help
to achieve the ultimate goal of obtaining a comprehensive and complete path
integral description of quantum mechanics and quantum field theory,
including quantum gravity and cosmology. \newline

\section*{Acknowledgments}

D-H Lin was supported by the National Youth Council of the ROC under
contract number NYC300375. Part of the work is supported by the rotary club
of Hsin Chu north west district.

\end{document}